\documentclass[12pt,A4widepaper]{article}
\usepackage{epsfig,amsmath,graphics}
\usepackage{graphics,graphicx}
\usepackage{tabularx}
\usepackage{amsfonts}
\usepackage{amsmath}
\usepackage{epstopdf}

\newcommand{\beq}{\begin{equation}}
\newcommand{\eeq}{\end{equation}}
\newcommand{\lb}{\label}
\newcommand{\beqar}{\begin{eqnarray}}
\newcommand{\eeqar}{\end{eqnarray}}
\newcommand{\barr}{\begin{array}}
\newcommand{\earr}{\end{array}}

\def\XXint#1#2#3{{\setbox0=\hbox{$#1{#2#3}{\int}$}
     \vcenter{\hbox{$#2#3$}}\kern-.5\wd0}}

\def\bD{\mbox{\boldmath${\it D}$}}

\def\bE{\mbox{\boldmath${\it E}$}}

\def\bF{\mbox{\boldmath${\it F}$}}

\def\b0{\mbox{\boldmath${\it 0}$}}

\def\bS{\mbox{\boldmath${\it S}$}}
\def\bs{\mbox{\boldmath${\it s}$}}

\def\bx{\mbox{\boldmath${\it x}$}}

\def\bchi{\mbox{\boldmath${\chi}$}}

\def\bLambda{\mbox{\footnotesize \boldmath${\Lambda}$}}

\def\grad{{\rm grad}}

\def\matR{\mbox{\boldmath${\mathbb R}$}}

\def\APL{{ Appl. Phys. Lett.\ }}

\def\IJSS{{ Int. J. Solids Structures\ }}

\def\JAP{{ J. Appl. Phys.\ }}

\def\SMS{{ Smart Mater. Struct.\ }}

\begin{document}

\title{\bf Optimal energy-harvesting cycles for load-driven dielectric generators in plane strain }

\author{R. Springhetti$^1$, E. Bortot$^1$, G. deBotton$^{1,2}$ and M. Gei$^1$\thanks{Corresponding author. 
Email: massimiliano.gei@unitn.it; web-page: www.ing.unitn.it/$\sim$mgei.} \\
   \\
{\sl \small $^1$Department of Civil, Environmental and Mechanical Engineering,}\\
{\sl \small University of Trento, Via Mesiano 77, I-38123 Trento, Italy;} \\
{\sl \small $^2$Department of Mechanical Engineering, Ben-Gurion University,}\\
{\sl \small  PO Box 653, Beer-Sheva 8410501, Israel.}}

\date{}
\maketitle
\vspace{-.5 cm}
\begin{center}
  {In honor of Professor Ray Ogden's 70th birthday}
\end{center}
\vspace{0.2 cm}

\begin{abstract}
The performances of energy harvesting generators based on dielectric elastomers are investigated.
The configuration is of a thin dielectric film coated by stretchable electrodes at both sides.
The film is first stretched, then charged and subsequently, afterwards it is released, and finally the charge is harvested at a higher electric potential.
The amount of energy extracted by this cycle is bounded by the electric breakdown and the ultimate stretch ratio of the film as well as by structural instabilities due to loss of tension.
To identify the optimal cycle that complies with these limits we formulate a constraint optimization problem and solve it with a dedicated solver for two typical classes of elastic dielectrics.
As anticipated, we find that the performance of the generator depends critically on the ultimate stretch ratio of the film.
However, more surprising is our finding of a universal limit on the dielectric strength of the film beyond which the optimal cycle is independent of this parameter.
Thus, we reveal that, regardless of how large the dielectric strength of the material is, there is an upper bound on the amount of harvested energy that depends only on the ultimate stretch ratio.
We conclude the work with detailed calculations of the optimal cycles for two commercially available elastic dielectrics.
\end{abstract}



\section{Introduction}
\label{sec:intro}

Among the various energy-harvesting technologies from renewable resources such as sea waves, wind, human gait and others, a particularly promising one is based on soft dielectric elastomers (DEs) \cite{AndersonReview,carpi2008,SRI2011,kornbluhpelrine_etc2011,mckayetal2011}.
A dielectric elastomer generator (DEG) is a highly deformable parallel-plate capacitor made out of a soft DE film coated with two compliant electrodes on its opposite faces.
The capacitance of the device depends on the deformation undergone by the DE film (through both the faces area and the thickness), hence changes during a load and release cycle resulting from an interaction of the device with its environment.
%
This variability can be exploited to extract electric energy by initially stretching, then charging the capacitor and subsequently releasing the capacitor and collecting the charge at a higher electric potential.




A few recent papers are dedicated to the analysis of the performances of DEGs and the identification of the more profitable electromechanical loading strategies during which the energy gain is maximized.
The contour of the region of admissible states which is dictated by typical failure modes of DEs was examined in \cite{KhoKeplingerBauerSuo} and \cite{kohzhaosuo2009} within the framework of finite electroelasticity.
A method to measure the produced energy and the efficiency of a balloon-like generator was developed in \cite{KaltseisKeplingerEtAl2011}.
In \cite{lall&etal12jpsb} various choices of electromechanical cycles for energy harvesting are presented.
An equibiaxial loading configuration of the device is assumed in \cite{suoclarkeadv2012}, who considered viscous effects too.
The possible benefits from embedding ceramic particles in a soft matrix on the amount of generated energy was considered in \cite{noi_cer2014}.

In the present work we focus on dielectric elastomer generators subjected to a four-stroke electromechanical cycle in which an external oscillating force powers the stretch and contraction cycle.
It is assumed that the device deforms under a plane-strain condition that simulates the effect of transverse constraint due to stiff fibers \cite{suoclarkesoftmatter2012} or a supporting frame.
Taking into account the properties of the elastic dielectric and the operating conditions dictated by the external environment, with the aid of a constrained optimization algorithm, our goal is to identify those cycles that produce the maximum energy.
This, in turn, will shade light on the relative role of the various failure mechanisms and provide guidelines for choosing  suitable elastic dielectrics for the DEG.
Time-dependent effects such as viscosity and loading frequencies are neglected and the results are determined assuming a conservative behaviour of the elastic dielectric.

The generator may undergo different failure mechanisms which must be avoided to ensure proper functioning and long service life of the device.
In the assumed plane-strain conditions the possible limits on the DEG performance are set by the electric breakdown threshold, a state of loss of tension with consequent buckling instability, and mechanical failure represented by the maximal allowable longitudinal stretch.
These failure limits, together with the assumption that the direction of the electric field is not reversed during the cycle, identify a contour of an admissible operational domain that can be depicted in the stress-stretch and the electric potential-charge planes.
In both representations the area enclosed within this contour is equal to a theoretical upper bound on the maximal energy that may be harvested from the DEG \cite{kohzhaosuo2009}.
The shape of the contour leads us in this work to distinguish between two types of optimal cycles depending on whether or not the electric breakdown limit is attained during the cycle.

We conclude the work with numerical analyses of two representative cases depending on the maximal allowable stretch for the cycle:
the first corresponds to a relatively stiff dielectric elastomer with a limited range of reversible stretches, whereas the second deals with a ductile elastomer that may be stretched a few times its referential length.
For both cases, general dimensionless analyses are carried out first for the two different types of optimal cycles.
Eventually, the analysis is specialized to two specific materials, a natural rubber and an acrylic elastomer, and the final results are given in a dimensional form.



\section{Theoretical background}
\label{sec:1}

\begin{figure}[!t]
\begin{center}
\includegraphics[width= 15 cm]{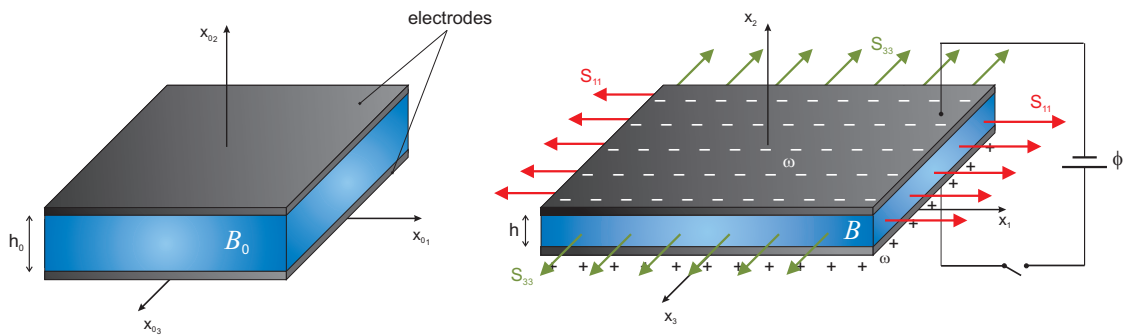}
\caption{\footnotesize {Reference and deformed configurations of a soft dielectric elastomer generator deforming in the  $\{x_1,x_2\}$-plane under plane-strain conditions.}}
\label{pb}
\end{center}
\end{figure}
%
In the course of this section we recall only those results that are pertinent to the subsequent analysis of DEGs.
These are based on the comprehensive analyses of dielectric layers subjected to electromechanical loading carried out in \cite{gdb&etal07mams,maxkatia2011,shmu&etal12ijnm}.
Essentially, a DEG is a stretchable capacitor, the basic idea behind its operating principle consisting in its ability to change the capacitance with deformation.
To clarify this concept, consider a dielectric elastomer occupying domains $B_{0}$ and $B\in \matR^3$ in the reference and the deformed configurations, respectively.
These are schematically illustrated in Fig.~\ref{pb}.
Throughout, cartesian coordinate systems are adopted in order to identify the positions of a point $\bx_0$ and $\bx = \bchi(\bx_0)$ in the two configurations, respectively.
Here, $\bchi$ is the {mapping} from the reference to the deformed configurations and $\bF=\partial{\bchi}/\partial{\bx_0}$ is the {deformation gradient}.

We consider an ideal dielectric film which is homogeneous, isotropic, hyperelastic, incompressible ($J\equiv\det\bF=1$), lossless and with no electrostrictive behaviour \cite{dorf&ogde05acmc,busta2009IJES,gei_electrostriction2014,mcmeeking,SuoEtAl2008}.
The film is stretched from the reference to the deformed configuration by a combination of ({\em i}) a mechanical force $\bs=[s,0,0]^T$ induced by the environment which is the primary source for the energy invested into the system,
and ({\it ii}) an electric field generated by an electric potential $\phi$ between the two stretchable electrodes coated on the opposite surfaces of the film at $x_2=0$ and $h$.
We note that an alternative way to electrically excite the deformation is by depositing electrical charge on the opposite surfaces of the specimen \cite{Kaltenbrunner}, however, in this work we do not consider this alternative since from a practical viewpoint it is more convenient to impose the required electric potential between the electrodes.

Neglecting fringing effects and assuming isotropy, 
the electromechanical deformation undergone by the film is homogeneous and can be represented by the deformation gradient $\bF={\rm diag}(\lambda, \lambda^{-1}, 1)$, where $\lambda$ is the principal stretch ratio along $x_1$.
Outside the capacitor the electric fields vanish, and the uniform electric field induced by the applied electric potential inside the film is $\bE=[0,E,0]^T$.
Recall that $\bE$ is conservative admitting the representation $\bE=-\grad \varphi(\bx)$, with $\varphi$ being the electrostatic potential field such that $\phi=\varphi|_{_{x_{2}=0}}-\varphi|_{_{x_{2}=h}}$, i.e. $E=\phi/h$.
In view of the homogeneous fields developing in the film, the applied force $s$ can be easily related to the nominal total stress $\bS$, which is divergence free when body forces are null.
Thus, along the prescribed loading path $\bS={\rm diag}(S_{11},S_{22},S_{33})$, where $S_{11}=s/h_0$, $S_{22}=0$ and $S_{33}$ is the reaction to the kinematic plane-strain constraint.
The energy conjugate to the electric field $\bE$ is the electric displacement field $\bD$, which is divergence free in the absence of free charges in the material.
Within the context of this work it is advantageous to represent the electric fields in terms of their referential counterparts $\bE_0=\bF^{T} \bE=[0,E_0,0]^T$, with $E_0=E/\lambda$, and $\bD_0=J \bF^{-1} \bD$.
In \cite{dorf&ogde05acmc} it is shown that the former is conservative and the latter is divergence free.

The electroactive hyperelastic incompressible material is assumed to be governed by an isotropic augmented energy-density function $W(\lambda_1,\lambda_2,\lambda_3,{E_0})$, where $\lambda_1$, $\lambda_2$ and $\lambda_3$ are the principal stretch ratios satisfying the incompressibility constraint $J=\lambda_1 \lambda_2 \lambda_3=1$.
The constitutive equations thus read
\beq
\lb{costeq}
S_{ii}=\frac{\partial W}{\partial \lambda_i}-p \frac{1}{\lambda_i},\ \ \ \ {D_0}=-\frac{\partial W}{\partial E_0},
\eeq
where $p$ is the unknown hydrostatic pressure, $D_0$ is the sole non-vanishing component of $\bD_0$, namely $\bD_0=[0,D_0,0]^T$, and no sum is implied on the components $S_{ii}$.

To illustrate the main features of the load-driven DEG, we adopt the simplest form of energy-density $W$, i.e. the one derived from neo-Hookean elasticity, namely
\beq
\lb{risposta specifico}
W=\frac{\mu}{2}(\lambda_1^2+\lambda_2^2+\lambda_3^2-3)-\frac{\epsilon}{2 }\left(\frac{{E_0}}{\lambda_2}\right)^2.
\eeq
Here $\mu$ is the shear modulus of the material and $\epsilon=\epsilon_r\epsilon_0$ its permittivity being $\epsilon_r$ the relative dielectric constant and $\epsilon_0=8.854$ pF/m the permittivity of vacuum.

In the present case $\lambda_1=\lambda,\lambda_2=\lambda^{-1}, \lambda_3=1$ and Eq.~(\ref{risposta specifico}) implies that $D_0=\epsilon\lambda^{2}\phi/h_{0}$.
Accordingly, the charge on the electrodes per unit of their undeformed area $\omega_0$ can be related to the stretch ratio $\lambda$ and the electric potential $\phi$ because the boundary condition on both sides corresponds to $\omega_{0}=D_0$.

In the sequel we find it advantageous to rephrase the equations in terms of the \emph{dimensionless} variables
\beq
\bar{S}_{11}= \frac{S_{11}}{\mu},\ \ \ \bar{S}_{33}= \frac{S_{33}}{\mu},\ \ \ \bar{\phi}= \frac{\phi}{h_0}\sqrt{\frac{\epsilon}{\mu}},\ \ \
\bar{\omega}_0=\frac{\omega_0}{\sqrt{\epsilon \mu}}.
\eeq
Next, with the aid of Eq.~(\ref{risposta specifico}) the components of the applied stress can be related to $\lambda$ and $\phi$.
Accordingly, during the harvesting cycle the relations between the applied stress, the applied electric potential, the resulting stretch ratio and the charge accumulated on the electrodes are:
\beq
\lb{dimless}
\bar{S}_{11}= \lambda - \frac{1}{\lambda^3} - {\bar{\phi}^2} \lambda,\ \ \
\bar{S}_{33}= 1 -  \frac{1}{\lambda^2} - {\bar{\phi}^2}\lambda^2,\ \ \
\bar{\phi}= \frac{\bar{\omega}_0}{\lambda^2}.
\eeq

\section{A load-driven generator and its failure envelope}
\label{sec:2}

\begin{figure}[!t]
\begin{center}
\includegraphics[width= 16 cm]{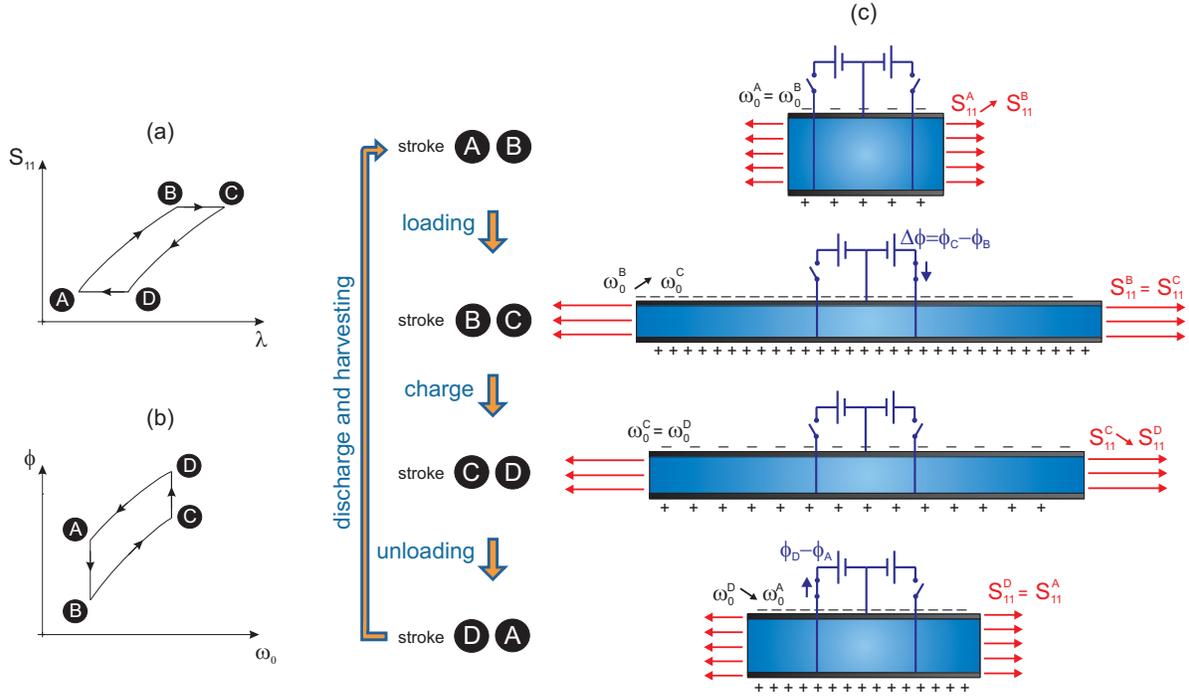}
\caption{\footnotesize {The harvesting cycle plotted on the mechanical plane (a) and the electrical plane (b), characterization of the four strokes with a service battery at the right and a storage battery at the left (c); all the plots in the descriptive sequence in (c) are referred to the initial state of the single stroke. }}
\label{cycle1}
\end{center}
\end{figure}
A few possible harvesting strategies, which are distinguished primarily according to the control parameters chosen for the four-stroke cycles, are discussed in \cite{lall&etal12jpsb}.
Here we focus on the cycle illustrated in Fig.~\ref{cycle1}:
along two of the strokes the longitudinal stress $S_{11}$ due to the external load is fixed, whereas along the other two strokes the device is electrically isolated and the charge $\omega_{0}$ in the electrodes is held fixed.
The order of these strokes and their role in the harvesting cycle are:
\begin{itemize}
\item a mechanical loading stroke A--B:
the work produced by the external oscillating force during its rise from minimal to maximal values is stored in the elastically stretching generator.
In terms of the dimensionless variables, $\bar{S}_{11}$ increases from its minimal value at A to its maximal value at B, while the charge on the electrodes is fixed ($\bar\omega_0^A=\bar\omega_0^B$).
Due to the stretching of the film the capacitance of the DEG increases and the electric potential drops;
%

\item an electrical charging stroke B--C: the electrodes are charged with the aid of an external service battery such that the electric potential between them is $\Delta\bar\phi=\bar\phi_C-\bar\phi_B>0$.
Along this step the stress is held constant ($\bar{S}_{11}^B=\bar{S}_{11}^C$).
Thanks to the attraction between the two charged electrodes the film further shrinks in the $x_{2}$-direction and elongates in the $x_{1}$-direction.
When this stroke terminates the film attains the largest stretch ratio during the cycle ($\lambda_C$);
%
%
\item a mechanical unloading stroke C--D: during the decline of the external force from its maximal to minimal values the film shrinks while the charge in the isolated electrodes is fixed ($\bar\omega_0^C=\bar\omega_0^D$).
The thickening of the shrinking film results in an increase of the electric potential to $\bar\phi_D$, which is the largest value of $\bar\phi$ along the cycle;
%

\item an electrical energy-harvesting stroke D--A: the charge deposited in stroke B--C is redeemed at a higher electric potential with an appropriate electrical circuit.
This energy-harvesting stroke is executed with a fixed load ($\bar{S}_{11}^D=\bar{S}_{11}^A$), nonetheless, due to the decrease in the electric potential between the electrodes and hence the associate decrease in the attracting force between them, the film further shrinks.
%
\end{itemize}
The cycle can be presented on the thermodynamical planes illustrated in Figs.~\ref{cycle1}(a) and (b), with
Fig.~\ref{cycle1}(a) corresponding to the mechanical $S_{11}$--$\lambda$ plane and Fig.~\ref{cycle1}(b) to the electrical $ \phi$--$\omega_0$ plane.
In passing, we note that in practice strokes B--C and D--A are substantially shorter than the mechanical loading and unloading strokes.
Thus, the applied external force should be thought of as a continuously oscillating force such that when it attains its maximal and minimal values appropriate electrical circuits are temporarily connected to the electrodes.

The four-stroke harvesting cycle described above is characterized by the four equalities presented in Fig. \ref{cycle1}(c).
These induce the following relations among the corresponding eight dimensionless independent variables:
\beqar
\lb{cyclerelations}
\bar{\omega}_0^A=\bar\omega_0^B\quad &\Rightarrow& \quad \bar{\phi}_A\lambda_A^2=\bar{\phi}_B\lambda_B^2,
\nonumber \\[2mm]
\bar S_{11}^B=\bar S_{11}^C \quad &\Rightarrow& \quad
\lambda_B - \frac{1}{\lambda_B^3} - {\bar{\phi}_B^2} \lambda_B= \lambda_C - \frac{1}{\lambda_C^3} - {\bar{\phi}_C^2} \lambda_C,  \nonumber
\\[2mm]
\bar\omega_0^C=\bar\omega_0^D\quad &\Rightarrow&  \quad  \bar{\phi}_C\lambda_C^2=\bar{\phi}_D\lambda_D^2,
\\[2mm]
\bar S_{11}^D=\bar S_{11}^A\quad   &\Rightarrow&  \quad
\lambda_D - \frac{1}{\lambda_D^3} - {\bar{\phi}_D^2} \lambda_D= \lambda_A - \frac{1}{\lambda_A^3} - {\bar{\phi}_A^2} \lambda_A.
\nonumber
\eeqar

In order to ensure a proper operational condition of the device, all feasible cycles must lie inside the region of admissible states for the generator.
The contour line that envelops this region is defined by the following possible failure modes of the DEG:
\begin{itemize}
\item electric breakdown (EB): this failure, which depends on the properties of the film, occurs when the electric field $E$ reaches the dielectric strength of the material $E_{eb}$.
In dimensionless form the dielectric strength is $\bar{E}_{eb}=E_{eb}\sqrt{\epsilon/\mu}$.
The dimensionless nominal electrical field is bounded by $\bar{E}_{0eb}=\bar{E}_{eb}/\lambda$.
The corresponding portions of the failure envelops surrounding the region of admissible states in the mechanical and the electric planes are prescribed in terms of the curves:
\beq
\lb{breakdown}
\bar{S}_{11} = \lambda - \frac{1}{\lambda^3} -\frac {\bar E_{eb}^2} {\lambda},\ \ \  \bar \phi=\frac{\bar E_{eb}^2}{\bar{\omega}_0};
\eeq

\item ultimate stretch ($\lambda_U$): this failure also depends on the properties of the film material and takes place when the magnitude of the stretch attains a critical value $\lambda_{U}$ at which mechanical failure initiates.
The curves that correspond to this failure mode in the mechanical and the electric planes are, respectively:
\beq
\lb{ultimate}
\lambda=\lambda_{U},\ \ \  \bar{\phi}=\frac{\bar{\omega}_0}{\lambda_{U}^2};
\eeq

\item Loss of tension ($S_{33}$=0): to avoid failure due to buckling instabilities it is required that the two in-plane stresses be positive.
In contrast with the previous two failure modes, this one is associated with the geometrical configuration of the device and is related to the small thickness of the film.
A comparison between expressions (\ref{dimless})$_{1}$ and (\ref{dimless})$_{2}$ for the two stresses reveals that the inequality $S_{33}\geq 0$ is more restrictive than $S_{11} \geq 0$.
Therefore, by manipulating these expressions it is found that the portions of the failure envelops corresponding to loss of tension along the $x_{3}$-direction in the two pertinent planes are characterized by the curves
\beq
\lb{stability}
\bar S_{11}=\lambda-\frac{1}{\lambda},\ \ \ \bar{\phi}=\frac{\bar{\omega}_0}{1+\bar{\omega}_0^2}.
\eeq
%
We stress that other types of instabilities may develop in the
electromechanically loaded film
\cite{rudy&etal12ijnm,puglisi_apl2013,gei_electrostriction2014}.
However, in this work only this type of instability is accounted for since
commonly electromechanical instabilities
do not play an important role in plane-strain conditions.
\end{itemize}
We finally add a fourth formal condition ($E$=0), which is not related to a failure of the mechanism, requiring that the direction of the electric field is not reversed during the cycle, i.e. $E\geq 0$.

\section{Optimization of the harvesting cycle}

According to the four-stroke cycle described in Section~\ref{sec:2}, the energy produced by the soft capacitor is
\beq
\lb{gain_total}
    \tilde H_g=\int_{B}^{C} \phi \, dQ + \int_{D}^{A} \phi \, dQ,
\eeq
where $Q$ is the total electric charge on the electrode.
Since the deformation is homogeneous, we prefer to work in dimensionless variables, introducing the energy-density generated per unit shear modulus
\beq
\lb{gain1}
 H_g=\frac{\tilde H_g}{\mu V_0}=\int_{B}^{C} \bar \phi \, d\bar\omega_0 + \int_{D}^{A}  \bar \phi \, d\bar\omega_0.
\eeq
The sum of the two integrals at the right-hand-side of Eq.~(\ref{gain1}) is equal to the area bounded within the cycle in the electrical plane in its dimensionless form (see Fig.~\ref{limiti} and the figures hereafter).
For later reference we recall that since $H_{g}$ is an energy which is extracted from the system its sign is negative.

Noting that $S_{11}$ is constant along both paths B--C and D--A, it is profitable to express the dimensionless electric potential $\bar \phi$ through Eq.~(\ref{dimless})$_1$ as a function of the stretch ratio $\lambda$ and the constant stress $S_{11}^{\,\rm const}=S_{11}^B$ and $S_{11}^{\,\rm const}=S_{11}^A$ along the two paths, respectively.
Accordingly,
\beq
\lb{phibarSconst}
 {\bar{\phi}} = \sqrt{1 - \lambda^{-4} - \lambda^{-1}\bar{S}^{\, \rm const}_{11}}.
\eeq
Moreover, Eq.~(\ref{dimless})$_3$ allows to evaluate the differential
\beq
\lb{diffomega0}
d\bar\omega_0=d\left[\bar{\phi}(\lambda) \lambda^{2}\right]=
\frac{4 \lambda -3 \bar{S}^{\, \rm const}_{11}}{2 \sqrt{{1 - \lambda^{-4} - \lambda^{-1}\bar{S}^{\, \rm const}_{11}}}}\,\, d\lambda.
\eeq
In turn, this enables the following explicit evaluation the first integral in (\ref{gain1}):
\beq
\lb{int1}
\int_{B}^{C} \bar \phi \, d\bar\omega_0 = \int_{\lambda_B}^{\lambda_C}\left(2 \lambda -\frac{3}{2} \bar{S}^{B}_{11}\right) d\lambda=
\left.\lambda\left(\lambda -\frac{3}{2} \bar{S}^{B}_{11}\right)\right|_{\lambda_B}^{\lambda_C}.
\eeq
The second integral is evaluated in a similar manner and successively added.
Next, the constant nominal stresses $\bar{S}^{B}_{11}$ and $\bar{S}^{A}_{11}$ are expressed in terms of the relevant stretches and dimensionless nominal electric potentials via Eq.~(\ref{dimless})$_1$.
This leads to the explicit expression for the dimensionless harvested energy:
\beq
\begin{split}
\lb{gain1calc}
 H_g=&\frac{1}{2}(\lambda_A-\lambda_D) \left[ \lambda_D\left(3 \bar\phi_D^2-1\right)+2 \lambda_A +3\lambda_D^{-3}\right]\\
&+\frac{1}{2}(\lambda_C-\lambda_B) \left[\lambda_B\left(3 \bar\phi_B^2-1\right) +2 \lambda_C +3\lambda_B^{-3}\right].
\end{split}
\eeq
Note that the expression for $H_{g}$ involves only the squares of the variables $\bar{\phi}_B$ and $\bar{\phi}_D$.
Therefore it is convenient to derive $\bar\phi_B^2$ and $\bar\phi_D^2$ from conditions (\ref{cyclerelations}) as functions of the stretches $\lambda_A$, $\lambda_B$, $\lambda_C$ and $\lambda_D$  and substitute them in (\ref{gain1calc}).
This will lead to an expression for $H_g$ in terms of the four characteristic stretches.
Similar developments can be followed for the constraints defining the failure envelope.
The final expressions determined for all the functions involved are outlined next.

%


\subsection{Constrained optimization problem}
\label{sec:constr_optim}

In order to determine the optimal cycle out of which the maximum energy can be harvested while keeping it within the region of admissible states, we formulate the following constrained optimization problem:
\beqar
\text{find}\,\min_{\bLambda}{H_g[\lambda_A,\lambda_B,\lambda_C,\lambda_D]} \nonumber
\eeqar
with $\bLambda=[\lambda_A,\lambda_B,\lambda_C,\lambda_D]^T$ and the minimum is sought since $H_{g}\le0$.
The optimization is to be evaluated under the following constraints:
\begin{itemize}
\item Equality constraint (active constraint)
\beq
\lb{eq_constrC}
f[\lambda_A,\lambda_B,\lambda_C,\lambda_D]=-\lambda_C+\lambda_U=0;
\nonumber
\eeq
The simplicity of this constraint enabled us to substitute $\lambda_{U}$ for $\lambda_{C}$ throughout the Lagrangian function and reduce the set of optimization variables $\bLambda$ to $\bLambda_{R}=[\lambda_A,\lambda_B,\lambda_D]^T$.

\item Inequality constraints (possibly active constraints)
\beqar
&&\!\!\!\!\!\!\!\!\!\!h_{1}[\lambda_A,\lambda_B,\lambda_C,\lambda_D]=S_{33}^D[\lambda_A,\lambda_B,\lambda_C,\lambda_D]\geq 0,\nonumber\\[2 mm]
&&\!\!\!\!\!\!\!\!\!\!h_{2}[\lambda_A,\lambda_B,\lambda_C,\lambda_D]=-\bar E_D^2[\lambda_A,\lambda_B,\lambda_C,\lambda_D]+\bar E_{eb}^2 \geq 0, \nonumber\\[2mm]
&&\!\!\!\!\!\!\!\!\!\!h_{3}[\lambda_A,\lambda_B,\lambda_C,\lambda_D]=\bar\phi_B^2 \ge 0\quad \text{i.e.}~\bar\phi_B\in \mathbb{R},\\[2mm]
&&\!\!\!\!\!\!\!\!\!\!h_{4}[\lambda_A,\lambda_B,\lambda_C,\lambda_D]=\lambda_A -1\ge 0,\qquad\quad
h_{5}[\lambda_A,\lambda_B,\lambda_C,\lambda_D]=-\lambda_A +\lambda_U\ge 0,\nonumber\\[2mm]
&&\!\!\!\!\!\!\!\!\!\!h_{6}[\lambda_A,\lambda_B,\lambda_C,\lambda_D]=\lambda_B -1\ge 0,\qquad\quad
h_{7}[\lambda_A,\lambda_B,\lambda_C,\lambda_D]=-\lambda_B +\lambda_U\ge 0,\nonumber\\[2mm]
&&\!\!\!\!\!\!\!\!\!\!h_{8}[\lambda_A,\lambda_B,\lambda_C,\lambda_D]=\lambda_D -1\ge 0,\qquad\quad
h_{9}[\lambda_A,\lambda_B,\lambda_C,\lambda_D]=-\lambda_D +\lambda_U\ge 0.\nonumber
\eeqar

\end{itemize}

The detailed expressions for the objective function $H_g$ and the contraints $h_1, h_2, h_3$ as functions of the stretches are:
\beqar
\lb{objective_fct}
H_g[\lambda_A,\lambda_B,\lambda_C,\lambda_D] \!\!&= &\!\! \frac{1}{2(\lambda_A^3 \lambda_C^3-\lambda_B^3 \lambda_D^3)}
                 \Big[-\lambda_C^3 \lambda_A^5 + 3 \lambda_D \left(\lambda_C^3-\lambda_D^2 \lambda_C+\lambda_B \lambda_D^2\right) \lambda_A^4 \nonumber\\[2 mm]
           & & -\left[3 (\lambda_B-\lambda_C)
                 \lambda_D^4+2 \lambda_C^3 \lambda_D^2+\lambda_C^3 \left(2 \lambda_B^2-3 \lambda_C \lambda_B+\lambda_C^2\right)\right] \lambda_A^3 \nonumber \\[2 mm]
           & & -2 \lambda_B^3 \lambda_D^3 \lambda_A^2 + 3 \lambda_B^3 \left[\lambda_D^4+(\lambda_B-\lambda_C) \lambda_C^3\right] \lambda_A  \nonumber \\[2 mm]
           & & -\lambda_B^3 \lambda_D \left[\lambda_D^4+(\lambda_B-2 \lambda_C)
                (\lambda_B-\lambda_C) \lambda_D^2+3 (\lambda_B-\lambda_C) \lambda_C^3\right]\Big],
\nonumber
\eeqar

\beqar
&&h_{1}[\lambda_A,\lambda_B,\lambda_C,\lambda_D]=\frac{\lambda_A^4 \lambda_C^3 \lambda_D-\lambda_A^3 \lambda_C^3 \left(\lambda_D^2-1\right)-\lambda_B^3 \lambda_D \left(\lambda_B \lambda_C^3-\lambda_C^4+\lambda_D^2\right)}
{\lambda_A^3 \lambda_C^3-\lambda_B^3 \lambda_D^3},\nonumber
\\[2mm]
&&h_{2}[\lambda_A,\lambda_B,\lambda_C,\lambda_D]=\frac{\lambda_A^4 \lambda_C^3 \lambda_D^3 - \lambda_A^3 \lambda_C^3 \left(\lambda_D^4-1\right) - \lambda_B^3 \lambda_D^3 \left(\lambda_B \lambda_C^3 - \lambda_C^4
+1\right)}{\lambda_D^2(\lambda_A^3 \lambda_C^3-\lambda_B^3 \lambda_D^3)}+\bar E_{eb}^2,\nonumber
\\[2mm]
&&h_{3}[\lambda_A,\lambda_B,\lambda_C,\lambda_D]=\frac{-\left(\lambda_A^4-1\right) \lambda_B^3 \lambda_D^3+\lambda_A^3 \lambda_B^3 \lambda_D^4+\lambda_A^3 \lambda_C^3 \left(\lambda_B^4-\lambda_B^3 \lambda_C-1\right)}
{\lambda_B^4(\lambda_A^3  \lambda_C^3-\lambda_B^3 \lambda_D^3)}.\nonumber
\eeqar

\noindent
The following generalized Lagrangian function is evaluated
\beq
\lb{lagrangian}
\begin{split}
\mathcal{L}[\lambda_A,\lambda_B,\lambda_D,\beta_1,\beta_2,\beta_3,\beta_4,\beta_5,\beta_6,\beta_7,\beta_8,\beta_9]&=\\
H_g[\lambda_A,\lambda_B,\lambda_U,\lambda_D]&
            - \sum_{i=1}^9 \beta_i\, h_{i}[\lambda_A,\lambda_B,\lambda_U,\lambda_D], \nonumber
\end{split}
\eeq
therefore, at every admissible (i.e. satisfying all the constraints) local minimum $\tilde\bLambda_{R}$, the following Karush--Kuhn--Tucker conditions are to be verified:
\beqar
\lb{lagrangian}
\left\{\begin{array}{lllll}
\nabla_{{\bLambda_{R}}} \mathcal{L}[\tilde\lambda_A,\tilde\lambda_B,\tilde\lambda_D,\tilde\beta_1,
\tilde\beta_2,\tilde\beta_3,\tilde\beta_4,\tilde\beta_5,\tilde\beta_6,\tilde\beta_7,\tilde\beta_8,\tilde\beta_9]=0,\\[2mm]
h_i[\tilde\lambda_A,\tilde\lambda_B,\lambda_U,\tilde\lambda_D]\ge0\quad\quad (i=1,\ldots, 9),\\[2mm]
\tilde\beta_i\ge 0\qquad\qquad\qquad\qquad\quad\!\! (i=1,\ldots,9),\\[2mm]
\tilde\beta_i\, h_i[\tilde\lambda_A,\tilde\lambda_B,\lambda_U,\tilde\lambda_D]=0\quad (i=1,\ldots,9).
\end{array} \right.
\nonumber
\eeqar
The gradient operator corresponds to the operation
\beqar
\nabla_{\bLambda_{R}}\mathcal{L}=
\left\{
\partial \mathcal{L}/\partial \lambda_A,\,\,\,
\partial \mathcal{L}/\partial \lambda_B,\,\,\,
\partial \mathcal{L}/\partial \lambda_D
\right\}^T,\nonumber
\eeqar
where the independent variables have been omitted for the sake of conciseness.
Finally, the stationarity condition reads:
\beq
\begin{split}
\nabla_{\bLambda_{R}}\mathcal{L}[\lambda_A,\lambda_B,\lambda_D,\beta_1,\beta_2,\beta_3,\beta_4,\beta_5,\beta_6,\beta_7,\beta_8,\beta_9]&=\\
\nabla_{\bLambda_{R}}H_g[\lambda_A,\lambda_B,\lambda_U,\lambda_D]\nonumber
            -&\sum_{i=1}^9 \beta_i\,\nabla_{\bLambda_{R}}h_{i}[\lambda_A,\lambda_B,\lambda_U,\lambda_D]=0. \nonumber
\end{split}
\eeq

\section{Numerical results}
In the spirit of the procedure described in Section~\ref{sec:constr_optim}, optimal energy-harvesting cycles have been determined with the aid of a dedicated numerical optimization procedure.
Specifically, the Nelder-Mead algorithm for constrained global optimization has been used in order to calculate optimal cycles for different values of the film parameters.
Results are illustrated hereafter for two choices of ultimate stretch ratios that are typical for commercial DEs.
However, before we proceed with the numerical results, it is useful to distinguish between two types of failure envelopes.

\subsection{Classification of possible failure envelopes}

\begin{figure}[!t]
\begin{center}
\includegraphics[width= 16 cm]{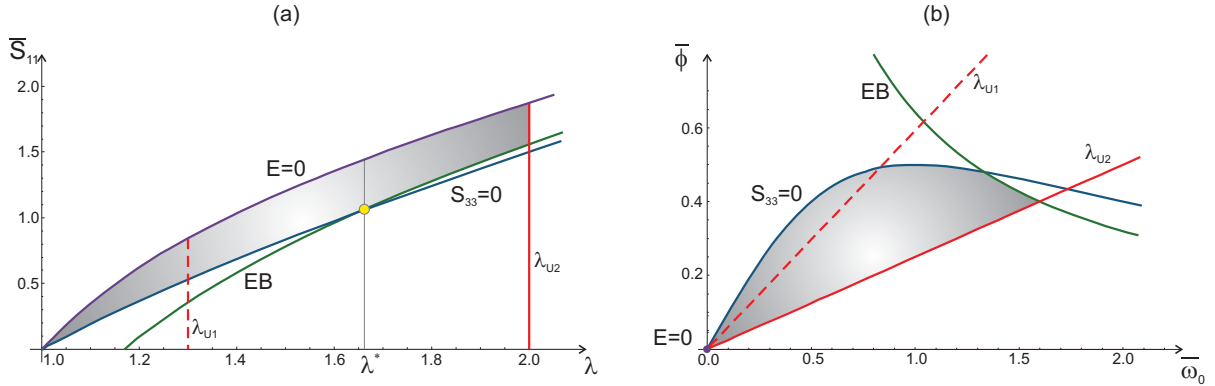}
\caption{\footnotesize {Regions of admissible states for DEG with $\bar E_{eb}=0.8$ and $\lambda^*=1.\bar{6}$ are shown on the mechanical (a) and the electrical (b) planes.
The dashed and continuous lines that represent the ($\lambda_U$) sections of the failure envelopes correspond to $\lambda_{U1}=1.3$ and $\lambda_{U2}=2$, respectively.}}
\label{limiti}
\end{center}
\end{figure}

A comparison between expressions (\ref{breakdown})$_{1}$ and  (\ref{stability})$_{1}$ for the stress according to the failure modes (EB) and ($S_{33}$=0) reveals the possible existence of an intersection point at which simultaneous failures may take place.
This is possible only if $\bar E_{eb}\leq 1$, and the combined failure occurs at stretch ratio $\lambda^*\equiv(1-\bar E_{eb}^2)^{-1/2}$ (see Fig.~\ref{limiti}).
If $\bar E_{eb}>1$ these two parts of the failure envelope do not intersect and the limit (EB) is not attainable by the DEG.
Because the failure modes (EB) and ($S_{33}$=0) may induce an independent limit on the stretch ratio of the film, we are led to the following distinction between two possible cases:
%

\begin{enumerate}
\item \lb{irrelevant EB} if $\lambda_U\leq\lambda^*$ the optimal cycle lies in the envelope dictated by ($S_{33}$=0), ($\lambda_U$) and ($E$=0) where the (EB) failure is unattainable;
\item \lb{relevant EB} if $\lambda_U>\lambda^*$ the failure mode (EB) must be accounted for in order to envelop the region of admissible states.
\end{enumerate}

\subsection{Moderately stretchable elastomers}
\label{1and5}

\begin{figure}[t]
\begin{center}
\includegraphics[width= 16 cm]{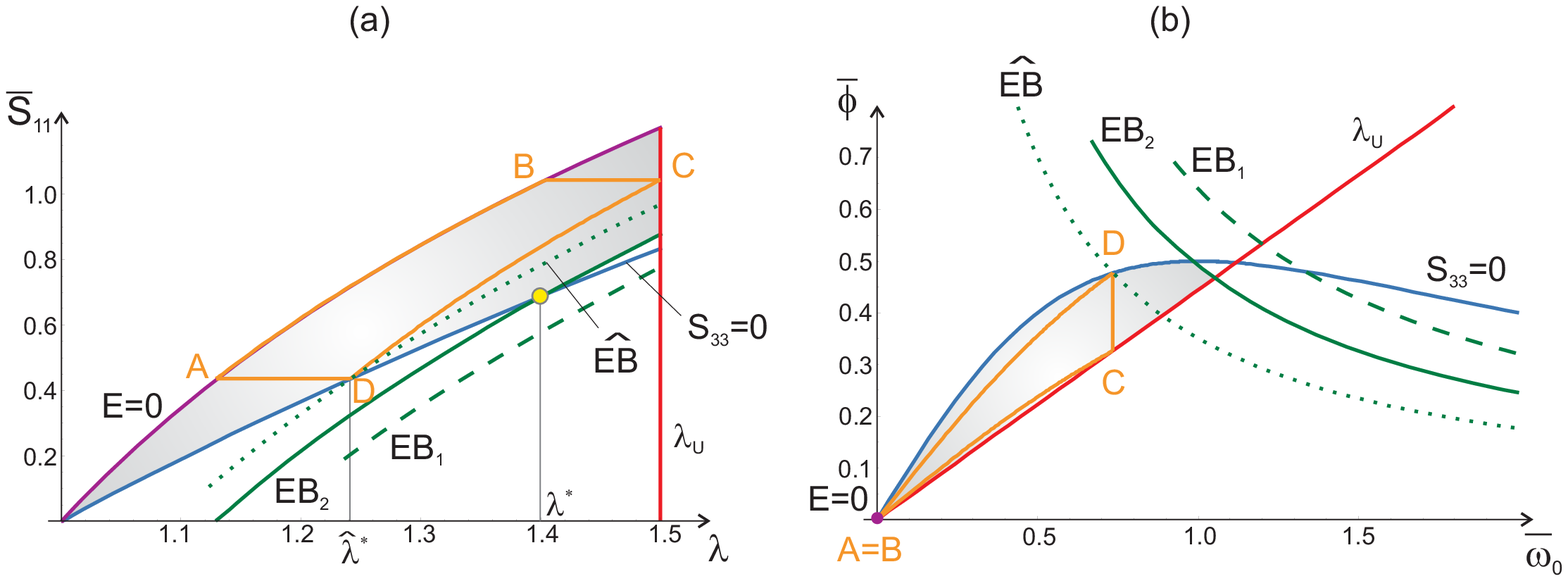}
\caption{\footnotesize {The optimal cycle for DE with $\lambda_U=1.5$ and $\bar E_{eb}\ge 0.5922$ plotted on the mechanical plane (a) and the electrical plane (b).
The  dashed and the continuous curves EB$_1$ and EB$_2$ correspond to $\bar E_{eb}=0.8$ (Case~\ref{irrelevant EB}) and $\bar E_{eb}=0.7$ (Case~\ref{relevant EB}a), respectively.
The dotted curve $\hat{\text{EB}}$ corresponds to the transition between Cases~\ref{relevant EB}a and \ref{relevant EB}b, at which ${\hat E}_{eb}=0.5922$.}}
\label{resL1e5Edown}
\end{center}
\end{figure}
\begin{figure}[t]
\begin{center}
\includegraphics[width= 16 cm]{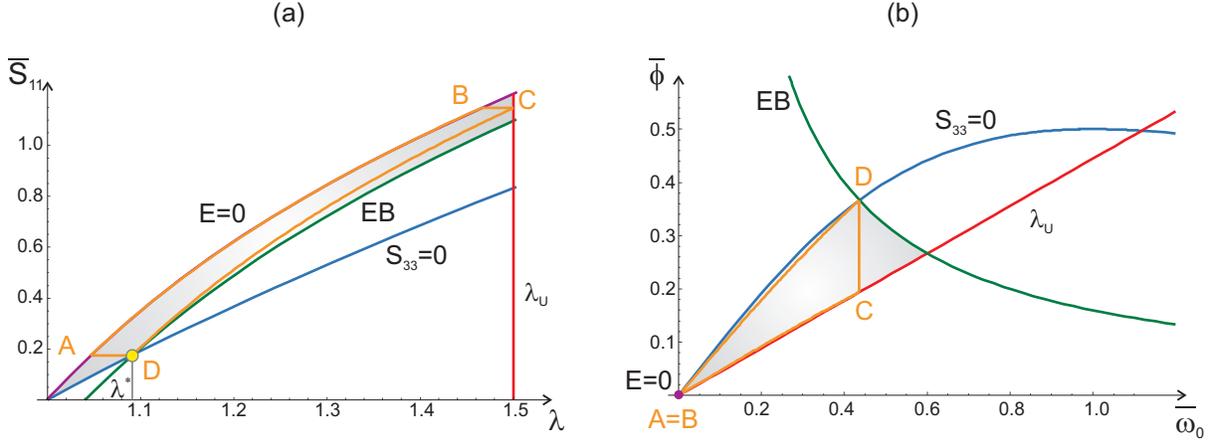}
\caption{\footnotesize {The optimal cycle for DE with $\lambda_U=1.5$ and $\bar E_{eb}=0.5<{\hat E}_{eb}$ (Case \ref{relevant EB}b) plotted on the mechanical plane (a) and the electrical plane (b).}}
\label{resL1e5E05}
\end{center}
\end{figure}

We consider first an elastic dielectric with a relatively small range of recoverable strains, $\lambda_{U}=1.5$.
For this relatively small stretch limit the failure curves ($S_{33}$=0) and (EB) do not intersect along the contour of the admissible domain as long as $\bar E_{eb}\geq0.7454$.
This corresponds to Case \ref{irrelevant EB} where (EB) is not attainable.
In Fig.~\ref{resL1e5Edown} the curve EB$_{1}$ that corresponds to $\bar E_{eb}=0.8$ demonstrates this case.
Smaller values of $\bar E_{eb}$ correspond to Case \ref{relevant EB}, for which we encounter the following two  occurrences:
\begin{itemize}

\item Case \ref{relevant EB}a: in the range ${\hat E}_{eb}\leq\bar E_{eb}<0.7454$ the optimal cycle is once again the one depicted in Fig.~\ref{resL1e5Edown}, where ${\hat E}_{eb}=0.5922$.
This sub-case is illustrated with the aid of the curve EB$_2$ that corresponds to $\bar E_{eb}=0.7$.
Thus, in a manner similar to the one observed for Case \ref{irrelevant EB}, at point D the optimal cycle is not limited by the dielectric strength of the material.

\item Case \ref{relevant EB}b: whenever $\bar E_{eb}<{\hat E}_{eb}$, the harvesting stroke of the optimal cycle starts at the point where the failure mechanisms ($S_{33}$=0) and (EB) are simultaneously activated, with $\lambda_D=\lambda^*$.
This sub-case is illustrated in Fig.~\ref{resL1e5E05} for $\bar E_{eb}=0.5$.
It is observed that in this case the optimal cycle is restricted by $\bar E_{eb}$.

\end{itemize}
Note that the transition between these two sub-cases at ${\hat E}_{eb}$ entails the coincidence of two points.
The first is the intersection point of the two boundaries of the failure envelope ($S_{33}$=0) and (EB), at which $\lambda=\lambda^*(\bar E_{eb})$.
The second point corresponds to point D of the cycle in the limit $\bar E_{eb}={\hat E}_{eb}$, where ${\hat E}_{eb}$ is the minimal value of the dielectric strength above which the failure curve (EB) is not attainable.
This transition is demonstrated in Fig.~\ref{resL1e5Edown} (for $\lambda_U=1.5$), where the dotted line represents the failure curve ($\hat{\text{EB}}$) for ${\hat E}_{eb}=0.5922$.
The stretch ratio at this point is $\hat\lambda^*=\lambda^*({\hat E}_{eb})=1.2410$.
%

Characteristic results for the optimal cycle determined for moderately stretchable DEs are summarized in Table \ref{tabledata1e5} for a few values of the dielectric strength.
Together with the amount of energy attained during the optimal cycle $|H_g|$, we also list the stretch ratio $\lambda^*$ at the intersection between the two failure curves (EB) and ($S_{33}$=0), the stretch ratio $\lambda_D$, the dimensionless activation electric potential $\Delta\bar\phi$ which is the potential supplied to the DEG during the electrical charging stroke B--C, and the anticipated failure mode at point D.
We note that for values of $\bar E_{eb}>{\hat E}_{eb}$ the maximal amount of harvested energy is fixed, while for lower values of $\bar E_{eb}$ the amount of extractable energy rapidly decreases.

%

\begin{table} [!h]
\centering
\begin{tabular}{ccccccc} \hline
[Ref.] &  $ \bar E_{eb}^{^{^{^{\,}}}}$&$\,\lambda^*$&$\,\lambda_D$&$|H_g|$&$\Delta\bar\phi^{^{\,}}$& Failure at D   \\
\hline
 Fig. \ref{resL1e5Edown} & $>0.7454$ & unessential &1.2410&0.0639& 0.3266& ($S_{33}$=$0$) \\
 Fig. \ref{resL1e5Edown} & 0.7 & 1.4003   &1.2410&0.0639& 0.3266& ($S_{33}$=$0$)\\
 Fig. \ref{resL1e5Edown} & 0.6 & 1.25     &1.2410&0.0639& 0.3266& ($S_{33}$=$0$)  \\
 Fig. \ref{resL1e5Edown} & $\hat{E}_{eb}=0.5922$ & 1.2410  &1.2410&0.0639& 0.3266& ($S_{33}$=$0$) \& (EB)  \\
 Fig. \ref{resL1e5E05}   & 0.5 & 1.1547   &1.1547&0.0562& 0.2566& ($S_{33}$=$0$) \& (EB) \\
 --                     & 0.4 & 1.0911   &1.0911 &0.0400& 0.1940& ($S_{33}$=$0$) \& (EB) \\
\hline
\end{tabular}
\caption{\footnotesize {Typical results for optimal cycles for DEs with $\lambda_U=1.5$ and different values of $\bar E_{eb}$.}}
\lb{tabledata1e5}
\end{table}

\subsection{Stretchable elastomers}

\label{3}

\begin{figure}[t]
\begin{center}
\includegraphics[width= 16 cm]{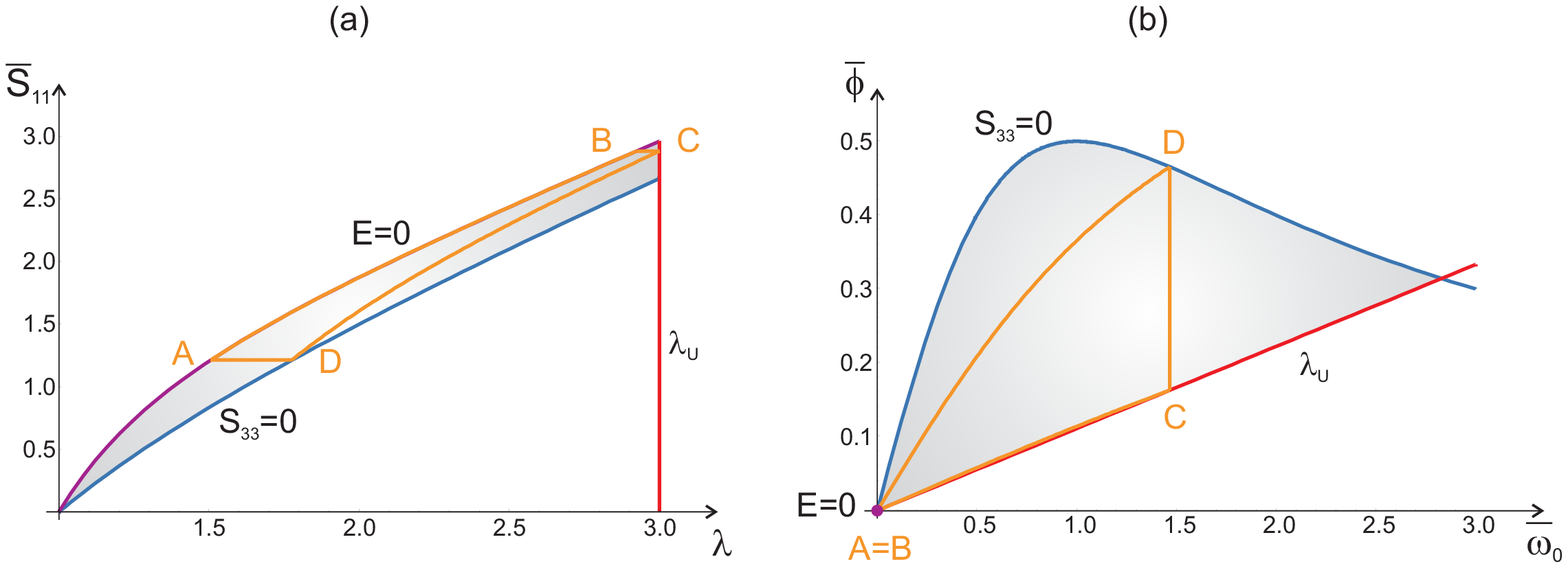}
\caption{\footnotesize {The optimal cycle for DEs with $\lambda_U=3$ and $\bar E_{eb}>{\hat E}_{eb}$ (Cases \ref{irrelevant EB} and \ref{relevant EB}a) plotted on the mechanical plane (a) and on the electrical plane (b).}}
\label{res3E1}
\end{center}
\end{figure}

\begin{figure}[t]
\begin{center}
\includegraphics[width= 16 cm]{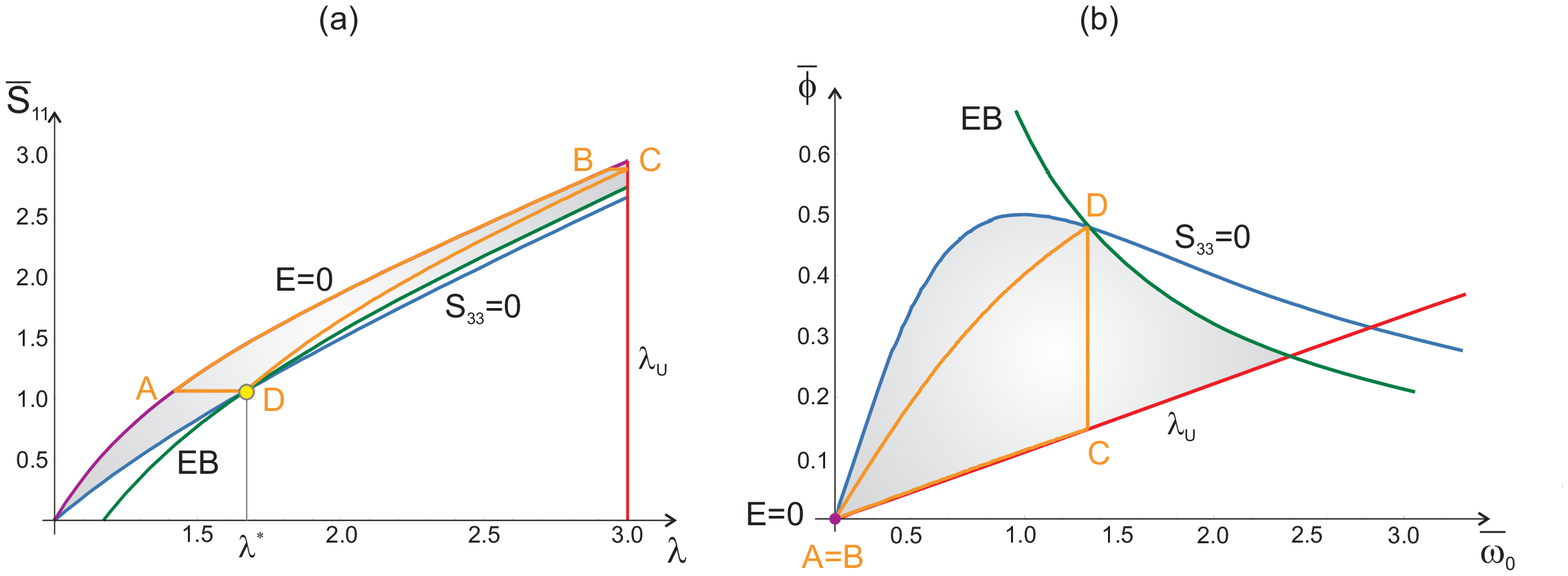}
\caption{\footnotesize {The optimal cycle for a DE with $\lambda_U=3$ and $\bar E_{eb}=0.8$ (Case \ref{relevant EB}b) plotted on the mechanical plane (a) and on the electrical plane (b).}}
\label{res3E08}
\end{center}
\end{figure}

\begin{figure}[!t]
\begin{center}
\includegraphics[width= 16 cm]{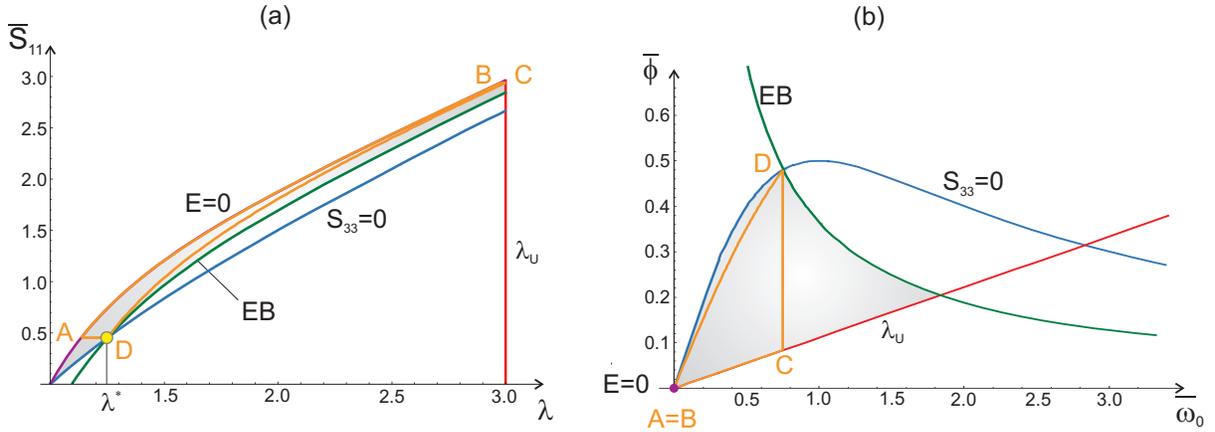}
\caption{\footnotesize {The optimal cycle for a DE with $\lambda_U=3$ and $\bar E_{eb}=0.6$ (Case \ref{relevant EB}b) plotted on the mechanical plane (a) and on the electrical plane (b).}}
\label{res3E06}
\end{center}
\end{figure}

\begin{table} [b]
\centering
\begin{tabular}{ccccccc} \hline
[Ref.] &  $ \bar E_{eb}^{^{^{^{\,}}}}$&$\,\lambda^*$&$\,\lambda_D$&$|H_g|$&$\Delta\bar\phi$& Failure at D   \\
\hline
 Fig. \ref{res3E1} & $>0.9428$ & unessential& 1.7756&0.2724 & 0.1630&  ($S_{33}$=$0$) \\
      --           & $\hat{E}_{eb}=0.8263$ & $1.7756$  & $1.7756$  &0.2724 &0.1630&  ($S_{33}$=$0$) \& (EB)\\
 Fig. \ref{res3E08} & 0.8 & $1.\bar 6$  & $1.\bar 6$  &0.2688 &0.1481&  ($S_{33}$=$0$) \& (EB)\\
 Fig. \ref{res3E06} & 0.6 & 1.25  & 1.25 &0.1658   &0.0833&  ($S_{33}$=$0$) \& (EB) \\
      --           & 0.5 & $1.1547$  & $1.1547$  &0.1146 &0.0642&  ($S_{33}$=$0$) \& (EB)\\
      --           & 0.4 & $1.0911$  & $1.0911$  &0.0727 &0.0485&  ($S_{33}$=$0$) \& (EB)\\
\hline
\end{tabular}
\caption{\footnotesize {Typical results for optimal cycles for DEs with $\lambda_U=3$ and different values of $\bar E_{eb}$.}}
\lb{tabledata3}
\end{table}
Results for stretchable DEs are depicted in Fig.~\ref{res3E1} for the cycle corresponding to Case~\ref{irrelevant EB} with $\bar E_{eb}\geq0.9428$.
In a manner similar to the one discussed for the moderately stretchable elastomers, Case~\ref{relevant EB} is further subdivided into Case~\ref{relevant EB}a for ${\hat E}_{eb}\leq\bar E_{eb} <0.9428$ and Case~\ref{relevant EB}b for smaller values of $\bar E_{eb}$.
The transition value of the dielectric strength for the stretchable DEs is ${\hat E}_{eb}=0.8263$.
In agreement with our previous discussion, as long as $\bar E_{eb} >{\hat E}_{eb}$ the optimal cycle which is shown in Fig.~\ref{res3E1} is unaffected by the failure curve (EB).
If the dielectric strength of the material is smaller than the transition value, the optimal cycle does depend on the failure curve (EB) as demonstrated in Figs.~\ref{res3E08} and \ref{res3E06} for $\bar E_{eb}=0.8$ and $0.6$, respectively.
A comparison between these two figures demonstrates the rapid decrease in the area bounded by the optimal cycles, corresponding to a severe drop in the harvested energy.
This observation is further evident from the summary presented in Table~\ref{tabledata3} for more values of $\bar E_{eb}$.
We observe that a two fold decrease in the dimensionless dielectric strength (0.83 $\rightarrow$ 0.4) results in almost a four fold decrease in the dimensionless harvested energy (0.27 $\rightarrow$ 0.073).
Yet, we stress again that for $\bar E_{eb} >{\hat E}_{eb}$, the amount of energy harvested is fixed.

\subsection{Material classification for DEG applications}
\begin{figure}[t]
\begin{center}
\includegraphics[width= 9 cm]{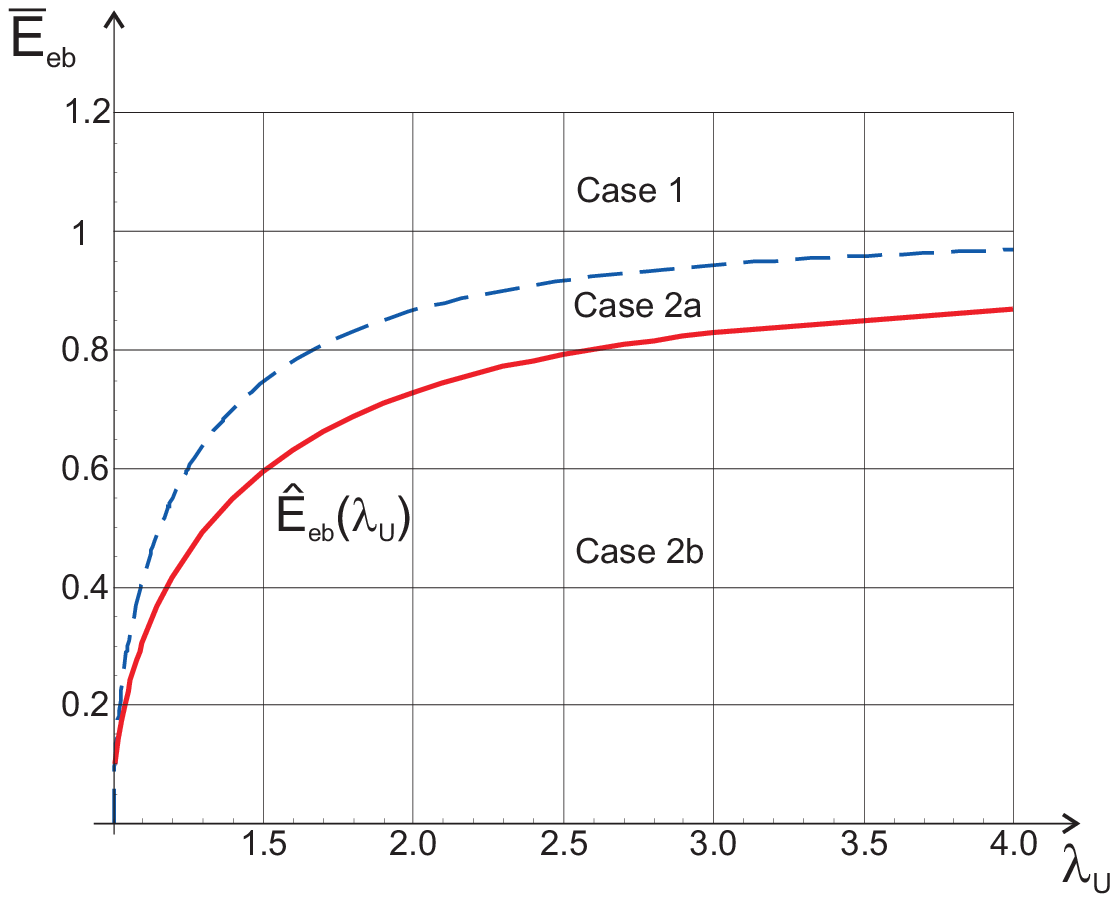}
\caption{\footnotesize {A universal curve (continuous) that distinguishes between two classes of DE film depicts the function $\hat E_{eb}(\lambda_{U})$. The dashed curve represents the transition between Cases 1 and 2.}}
\label{best_DE}
\end{center}
\end{figure}

\lb{materiale classificare}
The non-dimensional analyses carried out in subsection \ref{1and5} and  \ref{3} demonstrate that the amount of extractable energy from materials with large ultimate stretch ratio is larger.
This is evident on grounds of the extra work that may be produced by the external source thanks to the ability of a more stretchable DE to attain larger deformations.
A comparison between the results shown in Tables \ref{tabledata1e5} and \ref{tabledata3} further reinforce this conclusion.
Analogously, one may anticipate that the larger the dielectric strength of the DE is, the more energy could have been harvested from it.
This is because a larger electric field, and hence a higher electric potential gap between the electrodes, could be attained during the mechanical unloading stroke.
Here we find that while this is true at law values of dielectric strengths, at larger values there is a threshold beyond which the amount of harvested energy is fixed.
This, somewhat unanticipated finding, implies that in the context of this work there is a fine balance between the dielectric strength and the ultimate stretch ratio of the DE.
Thus, at one hand at low values of $E_{eb}$ the amount of extractable energy is low, and at the other hand larger values of $E_{eb}$ are unattainable.
In Fig.~\ref{best_DE} we highlight this result with the aid of a universal curve which is plotted on the $\lambda_{U}$--$\bar E_{eb}$ plane.
The continuous curve shows the variation of ${\hat E}_{eb}$ as a function of $\lambda_{U}$.
The optimal cycle of a DEG based on a dielectric whose pair of properties $\{\lambda_{U},\bar E_{eb}\}$ is located beneath this curve will depend on $\bar E_{eb}$ (Case \ref{relevant EB}b), and as was discussed before the amount of energy that may be harvested from this DEG is rather small.
If the property pair $\{\lambda_{U},\bar E_{eb}\}$ is located above this curve (Cases \ref{irrelevant EB} and \ref{relevant EB}a), the optimal cycle and hence the amount of harvested energy does not depend on $\bar E_{eb}$ and the DE full potentiality is not exploited.
Thus, in order to extract the maximum from the DEG it is recommended that its pair of properties $\{\lambda_{U},\bar E_{eb}\}$ be as close as possible to the universal curve shown in Fig.~\ref{best_DE}.

\subsection{DEGs based on commercially available materials}
We apply the analysis previously described to two specific commercially available DEs, the acrylic VHB-4910 and the acrylonitrile butadiene rubber (NBR).
The VHB-4910, produced by 3M, is a polyacrylate dielectric elastomer available as a pre-cast 1 mm thick polyacrylate adhesive foam.
The acrylonitrile butadiene rubber (NBR) is a synthetic elastomer, produced by co-polymerization of acrylonitrile (ACN) and butadiene rubber (BR).
This rubber exhibits enhanced actuation performance conferred by the presence of the ACN in the range between 18-50\% together with vulcanization agents \cite{Jung}.
The pertinent mechanical and electrical properties of the two materials listed in Table \ref{tabledimensional} have been reported in \cite{Jung}.

The dielectric strength of these materials, which commonly depends on the applied pre-strain, deserves a particular attention.
Thus, we recall that in common EAP applications the films are pre-strained in order to reduce the film thickness and increase the breakdown strength \cite{Kofod,TrolsKoglerEtAl2013}.
As an example, the data reported in \cite{Kofod} for VHB-4910 demonstrate that the electric breakdown limit ranges between 20 MV/m in the unstrained state and 218 MV/m in the case of 500\% equibiaxial strain.
Accordingly, in the sequel two typical values are assumed for the dielectric strength, ${E}_{eb_1}=20$ MV/m and ${E}_{eb_2}=100$ MV/m.
\begin{table} [!h]
\centering
\begin{tabular}{lccccc} \hline
Material  &[Ref.] & $\mu $ [kPa] & $\epsilon_r$ & $ \bar E_{eb_1}^{^{^{^{\,}}}}$ & $ \bar{E}_{eb_2}$  \\
\hline
VHB-4910& {\footnotesize \cite{Jung}}& 83.4   & 4.7 & 0.4468 & 2.2338 \\
NBR&      {\footnotesize \cite{Jung}}& 1333.4 & 14  & 0.1928 & 0.9642 \\
\hline
\end{tabular}
\caption{\footnotesize {Physical properties assumed for the elastic dielectrics.}}
\lb{tabledimensional}
\end{table}



To enable a comparison between the performances of load-driven DEGs based on the these materials, we took the VHB-4910 as a benchmark and determined the maximal stresses ($S_{\text{max}}=S_{11}^{B}$) attained along the optimal cycles corresponding to all four possible combinations of ${E}_{eb_1}$, ${E}_{eb_2}$, $\lambda_{U}=1.5$ and $\lambda_{U}=3$.
The four resulting stresses are listed in Table~\ref{tabledim_res2}.
We note that since the maximal stretch is attained at the end of the electric charging stroke (point C), the stresses corresponding to the larger dielectric breakdown are slightly lower since a larger portion of the allowable stretch is excited electrically.
Next, we determined limits on the maximal stretch $\lambda_{C}$ of the NBR such that, during the optimal cycles dictated by these stretches and the pair ${E}_{eb_1}$ and ${E}_{eb_2}$, the same maximal stresses will be developed.
The resulting stretch ratios for the NBR, together with the harvested energy and the normalized activation potential, are also listed in Table~\ref{tabledim_res2}.
In this table we also list these materials classification (M-C) according to Fig.~\ref{best_DE}.
In agreement with the one order of magnitude larger shear modulus of the NBR, the respective stretches that correspond to similar maximal stresses are an order of magnitude smaller for the NBR.

\begin{table} [!h]
\centering
{\small
\begin{tabular}{l|ccccc|ccccc}\hline
&\multicolumn{5}{c}{$ E_{eb_1}=20$ MV/m} &\multicolumn{5}{|c}{$ E_{eb_2}=100$ MV/m}\\
 \hline
             & $S_{\text{max}} $ & $\mu H_{g} $ & $\Delta\phi/h_0$ & $\lambda_{C}$ & M-C & $S_{\text{max}} $ & $\mu H_{g} $& $\Delta\phi/h_0$ & $\lambda_{C}$ & M-C\\
  Material        &\!\![kPa]\!\! &\!\![kJ/m$^{3}$]\!\! & \!\!\! \!\!   [kV/mm] \!\!\!\!\!     &  &  &\!\![kPa]\!\! &\!\![kJ/m$^{3}$]\!\!  &\!\!\!\!\!    [kV/mm] \!\!\!\!\!  &  &  \\
\hline
VHB-4910 &94.2 & 3.99 &9.9 & 1.5 & (2b)  &87.0 & 5.33 & 14.62 & 1.5 & (1)  \\
NBR   &94.2 & 0.37  &15.4 &1.024 & (2a)  &87.0 & 0.32 &14.81 & 1.022  & (1) \\
\hline
VHB-4910 &246.3 & 7.60 &2.5 &3 & (2b)  &240.5 & 22.72 & 7.3 &3 &  (1) \\
NBR   &246.3 & 1.86 &18.2 &1.059  & (2b) &240.5 & 2.48 &23.3 &1.064 & (1)  \\
\hline
\end{tabular}
}
\caption{\footnotesize {The harvested energy-density $\mu H_g$, activation electric potential per referential unit thickness $\Delta\phi/h_0$ and material classification M-C, determined for the optimal cycles according to the dielectric strength limits and the maximal stretch ratios $\lambda_{C}$ determined for the maximum stresses $S_\text{max}$.}}

\lb{tabledim_res2}
\end{table}

At first glimpse we deduce from Table~\ref{tabledim_res2} that the amounts of energy extractable from the VHB-4910 are an order of magnitude larger than those available for the NBR (when the load or the traction on the DEG are specified).
This is due to the extensibility of the former and in spite of the fact that the dielectric constant of the VHB-4910 is substantially smaller than that of the NBR.
The critical impact of the stretchability of the film can be further observed when comparing the results determined for the NBR in the second line of Table~\ref{tabledim_res2}.
We note that there is a small difference between the maximal stretch ratios $\lambda_{C}$ in favor of the material with the lower dielectric breakdown.
Nonetheless, this small difference is responsible for approximately 20\% more energy that can be produced with a DEG based on the NBR with smaller $E_{eb}$.

Another important observation concerns the importance of the classification of the materials.
In particular we note the differences in the energies harvested from the VHB-4910.
In the first line of Table~\ref{tabledim_res2} the energy associated with the VHB-4910 for smaller dielectric strength is comparable with the one extracted from the material whose dielectric strength is 5 times larger.
This is because the material with larger dielectric breakdown is associated with Case~\ref{irrelevant EB}, according to which this larger dielectric strength is not reached during the optimal cycle, so that the potential of the material remains unexpressed.
Along the third line of Table~\ref{tabledim_res2}, where VHB-4910 with larger stretchable domain is considered the situation is somewhat different:
the energy extracted from the material with larger $E_{eb}$ is approximately three times the one extracted from the one with smaller $E_{eb}$.
This is because the pair $\{\lambda_{U},\bar E_{eb}\}$ for the material with larger dielectric breakdown is closer to the universal curve shown in Fig.~\ref{best_DE}, suggesting that this material is closer to the optimal balance between the ultimate stretch ratio and the dielectric breakdown.

\section{Conclusions}
We investigated the performances of load-driven DE generators made out of a dielectric film coated with compliant electrodes at both sides, undergoing electromechanical Carnot-type cycles in a plane-strain condition.
The cycle assumed in our study is composed of the following four strokes:
({\em i}) through-thickness thinning and longitudinal stretching driven by an increasing external force at constant electric charge;
({\em ii}) application of an electric potential under constant load allowing charge deposition onto the electrodes;
({\em iii}) longitudinal contraction under a decreasing mechanical load and constant electric charge, with the electric potential increasing correspondingly;
({\em iv}) discharge of useable charge at a constant load.
Overall a net amount of electrostatic energy is released over a single cycle.
In order to identify the best cycle out of which maximum energy can be harvested, a constraint optimization problem was  formulated accounting for possible failure modes of the DEG.
The failure modes accounted for are electric breakdown, mechanical rupture due to an over stretching of the film and the development of instabilities due to loss of tension.
These were characterized in terms of appropriate limits on the electrical field, the maximal stretch ratio and the stresses developing in the film.

The constraint optimization problem was solved with the aid of a dedicated algorithm for two typical classes of elastomers.
As anticipated, the performance of a DEG crucially depends on the stretchability of the elastic dielectric.
This is because the mechanical work produced by the external force can be larger if the film can reach large reversible stretch ratios.
An important and less trivial consequence of the conducted analysis concerns the relationship between the attainable stretch ratio and the dielectric strength of the material.
This is presented in terms of the universal curve shown in Fig.~\ref{best_DE}.
We find that for materials with \lq low' dielectric strength the optimal cycle attains the electric breakdown field.
The area beneath the solid curve in Fig.~\ref{best_DE} corresponds to this class of materials.
Clearly, the larger the dielectric strength of materials in this regime is, the more energy can be harvested from the DEG.
However, if the dielectric strength of the film is larger than the threshold limit represented by this curve, the maximal electric field that develops during the optimal cycle does not reach the electric breakdown of the material.
This implies that if the pair dielectric strength--ultimate stretch ratio of the film corresponds to a point above the universal curve, the maximal harvested energy is independent of the dielectric strength of the film.
In other words, once the ultimate stretch ratio of the film is given, the maximal extractable energy is bounded by the universal curve of Fig.~\ref{best_DE} regardless of the material electric breakdown limit.
The universal curve may also serve  as a design criterion for DE-based generators.
Thus, in order to obtain the maximum energy from a given material at one hand and to use the full capacity of this material at the other, the most beneficial working cycle of the DEG is the one in which the initiation of the harvesting cycle lays as close as possible to this curve.



To highlight the results of our analysis we concluded this work with a comparison between the performances of generators based on two commercially available DEs, the VHB-4910 and the NBR.
The stretch limits for these materials were set in such a way that the load driving the DEGs based on the two materials is identical.
We find that thanks to the large stretches undergone by the VHB-4910 the amount of energy that can be harvested from DEGs based on this material is larger than the amount produced by the NBR.
We finally demonstrate that DEGs that are based on films whose dielectric breakdown and ultimate stretch are close to the universal curve are more efficient in the sense that these films can be used to almost their full potential.


\section*{Acknowledgements}

GdB gratefully acknowledges the 2013 Visiting Professorship Program of the University of Trento.
GdB also acknowledges ESNAM (European Scientific Network for Artificial Muscles) COST Action MP1003 for supporting STSM no. 12316 during which the reported study was initiated.
MG acknowledges the support of PRIN grant no. 2009XWLFKW, financed by Italian Ministry of Education, University and Research.


\begin{small}

\end{small}


\begin{thebibliography}{99}


\bibitem{AndersonReview}
 I.A. Anderson, T.A. Gisby, T.G. McKay, B.M. O'Brien, E.P. Calius
 {\em Multi-functional dielectric elastomer artificial muscles for soft and smart machines}.
 \JAP 112, 041101, 2012.


\bibitem{maxkatia2011}
 K. Bertoldi, M. Gei
 {\em Instability in multilayered soft dielectrics}.
 J. Mech. Phys. Solids 59, 18--42, 2011.

\bibitem{noi_cer2014}
 E. Bortot, R. Springhetti, M. Gei
 {\em Enhanced soft dielectric composite generators: the role of ceramic fillers}.
 J. Eur. Ceram. Soc., in press, 2014.

\bibitem{busta2009IJES}
R. Bustamante, A. Dorfmann, R.W. Ogden
{\em On electric body forces and Maxwell stresses in nonlinearly electroelastic solids}.
Int. J. Engin. Science, 47, 1131--1141, 2009.

\bibitem{carpi2008}
 F. Carpi, D. De Rossi, R. Kornbluh, R. Pelrine, P. Sommers-Larsen (Eds). {\em Dielectric elastomers as electromechanical transducers --
 Fundamentals, Materials, Devices, Models and Applications of an Emerging Electroactive Polymer Technology}.
 Elsevier, Oxford, UK, 2008.

\bibitem{SRI2011}
 S. Chiba, M. Waki, R. Kornbluh, R. Pelrine {\em Current status and future prospects of power generators using dielectric elastomers}.
 Smart Materials and Structures 20, 124006, 2011.

\bibitem{gdb&etal07mams}
G.~de{B}otton, L.~Tevet-Deree, and E.A. Socolsky
{\em Electroactive heterogeneous polymers: analysis and applications to laminated composites}.
Mechanics of Advanced Materials and Structures 14, 13--22, 2007.

\bibitem{puglisi_apl2013}
 D. De Tommasi, G. Puglisi, G. Zurlo
 {\em Electromechanical instability and oscillating deformations in electroactive polymer films}.
 \APL 102, 011903, 2013.

\bibitem{dorf&ogde05acmc}
 A. Dorfmann, R.W. Ogden,
 {\em Nonlinear electroelasticity}.
 Acta Mech. 174, 167--183, 2005.

\bibitem{gei_electrostriction2014}
 M. Gei, S. Colonnelli, R. Springhetti
 {\em The role of electrostriction on the stability of dielectric elastomer actuators}.
 \IJSS 51, 848--860, 2014.


\bibitem{suoclarkeadv2012}
 J. Huang, S. Shian, Z. Suo, D.R. Clarke
 {\em Maximizing the Energy Density of Dielectric Elastomer
 Generators Using Equi-Biaxial Loading}.
 Adv. Funct. Mat. 1--6, 2013

\bibitem{Jung}
 K. Jung, J. Lee, M. Cho, J.C. Koo, J. Nam, Y. Lee, H.R. Choi
 {\em Development of enhanced synthetic elastomer for energy-efficient polymer actuators}.
 \SMS 16, S288--S294, 2007.


\bibitem{KaltseisKeplingerEtAl2011}
 R. Kaltseis, C. Keplinger, R. Baumgartner, M. Kaltenbrunner, T. Li, P. M\"achler, R. Schw\"odiauer, Z. Suo, S. Bauer
 {\em Method for measuring energy generation and efficiency of dielectric elastomer generators}.
 \APL 99, 162904, 2011.

\bibitem{Kaltenbrunner}
C. Keplinger, M. Kaltenbrunner, N. Arnold and S. Bauer.
Rontgen's electrode--free elastomer actuators
without electromechanical pull--in instability.
PNAS 107, 4505--4510, 2010.

\bibitem{Kofod}
 G. Kofod, R. Kornbluh, R. Pelrine, P.Sommer-Larsen
 {\em Actuation response of polyacrylate dielectric elastomers}.
 J. Intell. Mater. Syst. and Struct. 14, 787--793, 2003.

\bibitem{KhoKeplingerBauerSuo}
 S.J.A. Koh, C. Keplinger, T. Li, S. Bauer, Z. Suo
 {\em Dielectric elastomer generators: how much energy can be converted?}
 IEEE/ASME Trans. on Mechatronics 16, 1083--4435, 2011.

\bibitem{kohzhaosuo2009}
 S.J.A. Koh, X. Zhao, Z. Suo  {\em Maximal energy that can be converted by a dielectric elastomer generator}.
 \APL 94,  262902, 2009.

\bibitem{kornbluhpelrine_etc2011}
 R.D. Kornbluh, R. Pelrine, H. Prahlad, A. Wong-Foy, B. McCoy, S. Kim, J. Eckerle, T. Low  {\em From boots to buoys: promises and challenges of dielectric elastomer energy harvesting}.
 In Y.B. Cohen and F. Carpi (Eds),  {\em Electroactive polymer actuators and devices (EAPAD)}, Vol. 7976, Bellingham, WA, 2011.

\bibitem{lall&etal12jpsb}
M. Lallart, P.J. Cottinet, D. Guyomar, L. Lebrun
{\em Electrostrictive polymers for mechanical energy harvesting}.
Journal of Polymer Science part B: Polymer Physics 50, 523--535, 2012

\bibitem{suoclarkesoftmatter2012}
 T. Lu, J. Huang, C. Jordi, G. Kovacs, R. Huang, D.R. Clarke, Z. Suo
 {\em Dielectric elastomer actuators under equal-biaxial forces, uniaxial forces, and
 uniaxial constraint of stiff fibers}.
 Soft Matter 8, 6167--6173, 2012.

\bibitem{mckayetal2011}
 T.G. McKay, B.M. O'Brien, E.P. Calius, I.A. Anderson  {\em Soft generators using dielectric elastomers}.
\APL 98,  142903, 2011.

\bibitem{mcmeeking}
 R.M. Mc{M}eeking, C.M. Landis
 {\em Electrostatic forces and stored energy for deformable dielectric materials}. J. Appl. Mech. 72, 581--590, 2005.

\bibitem{rudy&etal12ijnm}
S.~Rudykh, K.~Bhattacharya, and G.~de{B}otton.
{\em Snap-through actuation of thick-wall electroactive balloons}.
Int. J. Nonlinear Mech. 47, 206--209, 2012.


\bibitem{shmu&etal12ijnm}
G.~Shmuel, M.~Gei, and G.~de{B}otton.
{\em The {R}ayleigh-{L}amb wave propagation in dielectric elastomer layers subjected to large deformations}.
Int. J. Nonlinear Mech., 47, 307--316, 2012.

\bibitem{SuoEtAl2008}
 Z. Suo, X. Zhao, W.H. Green
 {\em A nonlinear field theory of deformable dielectrics}.
 J. Mech. Phys. Solids 56, 467--486, 2008.

\bibitem{TrolsKoglerEtAl2013}
 A. Tr\"ols, A. Kogler, R. Baumgartner, R. Kaltseis, C. Keplinger, R. Schw\"odiauer, I. Graz, S. Bauer
 {\em Stretch dependence of the electrical breakdown strength and dielectric constant of dielectric elastomers}.
 Smart Mat. Struct. 22, 104012, 2013.









\end{thebibliography}
\end{document}